\documentclass[prd,showpacs,preprintnumbers,amsmath,amssymb]{revtex4}

\usepackage{graphicx}
\usepackage{dcolumn}
\usepackage{bm}

\usepackage{color}

\begin{document}

\title{Schwinger Effect in Near-extremal Charged Black Holes in High Dimensions}

\author{Rong-Gen Cai}\email{cairg@itp.cas.cn}
\affiliation{CAS Key Laboratory of Theoretical Physics, Institute of Theoretical Physics, Chinese Academy of Sciences, P.O. Box 2735, Beijing 100190, China}
\affiliation{School of Physical Sciences, University of Chinese Academy of Sciences, No.19A Yuquan Road, Beijing 100049, China}

\author{Chiang-Mei Chen} \email{cmchen@phy.ncu.edu.tw}
\affiliation{Department of Physics, National Central University, Chungli 32001, Taiwan}
\affiliation{Center for High Energy and High Field Physics (CHiP), National Central University, Chungli 32001, Taiwan}

\author{Sang Pyo Kim}\email{sangkim@kunsan.ac.kr}
\affiliation{Department of Physics, Kunsan National University, Kunsan 54150, Korea}
\affiliation{Institute of Theoretical Physics, Chinese Academy of Sciences, Beijing 100190, China}

\author{Jia-Rui Sun}\email{sunjiarui@mail.sysu.edu.cn}
\affiliation{School of Physics and Astronomy, Sun Yat-Sen University, Guangzhou 510275, China}

\date{\today}

\begin{abstract}
We study the Schwinger effect in near-extremal nonrotating black holes in an arbitrary $D(\geq 4)$-dimensional asymptotically flat and (A)dS space. Using the near-horizon geometry $\mathrm{AdS}_2 \times \mathrm{S}^{D-2}$ of near-extremal black holes with Myers-Perry metric, we find a universal expression of the emission formula for charges that is a multiplication of the Schwinger effects in an $\mathrm{AdS}_2$ space and in a two-dimensional Rindler space. The effective temperature of an accelerated charge for the Schwinger effect is determined by the radii of the effective $\mathrm{AdS}_2$ space and $\mathrm{S}^{D-2}$ as well as the mass, charge, angular momentum of the charge and the radius of the (A)dS space. The Schwinger effect in the asymptotically flat space is more efficient and persistent for a wide range of large black holes for dimensions higher than four. The AdS (dS) boundary enhances (suppresses) the Schwinger effect than the asymptotically flat space. The Schwinger effect persists for a wide range of black holes in the AdS space and has an upper bound in the dS space.
\end{abstract}


\maketitle

\section{Introduction}

The charged black hole provides an interesting model to explore quantum nature of black holes because of Hawking radiation~\cite{Hawking:1974sw} and the Schwinger pair production of charges from the electric field of the black hole. The extremal charged black hole, in particular, has the zero-Hawking temperature but its electric field can create charged particles through the Schwinger effect of pair production~\cite{Schwinger:1951nm}. The near-horizon geometry of $\mathrm{AdS}_2 \times S^2$ of four-dimensional (near-) extremal Reissner-Nordstr\"{o}m (RN) black holes has allowed one to obtain an explicit formula for the Schwinger effect~\cite{Chen:2012zn, Chen:2014yfa}. Near-extremal Kerr-Newman black holes with electric and/or magnetic charges have a warped AdS geometry, whose emission formula is also found~\cite{Chen:2016caa, Chen:2017mnm}. The Schwinger effect in the $\mathrm{AdS}_2$ sector is governed by the effective temperature for accelerated charges by the electric field on the horizon~\cite{Cai:2014qba}. One interesting aspect is that the Hawking radiation and the Schwinger effect are intertwined for near-extremal black holes, which may shed a light on understanding radiation from charged black holes beyond the Hawking radiation. (For review and references, see Ref.~\cite{Kim:2019joy}.)

In this paper we study the Schwinger effect in (near-) extremal nonrotating charged black holes in an arbitrary $D (\geq 4)$-dimensional,  asymptotically flat, AdS or dS spacetime. The Einstein-Maxwell theory in a $D$-dimensional spacetime has charged black holes with the Myers-Perry (MP) metric~\cite{Myers:1986un}. The near-extremal black holes have the near-horizon geometry of $\mathrm{AdS}_2 \times S^{D-2}$, in which a probe charged field can be solved in terms of hypergeometric or Whittaker functions. In contrast to the four-dimensional RN black hole, the charge differently feels the effective radii for the $\mathrm{AdS}_2$ and $\mathrm{S}^{D-2}$ in higher dimensions. The Schwinger effect exhibits this property in the four-dimensional (A)dS space~\cite{Chen:2020mqs}. Recently the holographic description of the Schwinger effect in five-dimensional RN brane in the $\mathrm{AdS}$ space has been given in Ref.~\cite{Zhang:2020apg}.

It is shown that the Schwinger pair production from near-extremal charged black holes, regardless of the asymptotically flat or (A)dS spacetime, has a universal formula which is factorized into the Schwinger formula governed by an effective temperature for accelerated charge by an electric field in $\mathrm{AdS}_2$ and another Schwinger formula for accelerated charge in the two-dimensional Rindler space with the Hawking temperature. Remarkably, this factorization seems to be universal and the Schwinger effect and the Hawking radiation cannot be separated, though one effect may dominate the other depending on the ratio of charge to mass of the black hole. The effect of the dimensionality and the (A)dS radius on the Schwinger effect is investigated. An enhancement of the Schwinger effect due to high dimensions is observed in the asymptotically flat spacetime
and the (A)dS boundary effect on the Schwinger pair production persists for a wide range of black hole radius.

The organization of this paper as follows. In Sec.~\ref{MP sec}, the near-horizon geometry of MP metric is shown to have $\mathrm{AdS}_2 \times S^{D-2}$. We find the solutions of charged scalar field in the near-horizon region. In Sec.~\ref{Sch sec}, we find the mean number of emitted charges and show the universality of the formula in near-extremal charged black holes. In Sec.~\ref{CC sec}, we study the effect of the asymptotic boundary of (A)dS on the Schwinger effect. In $D=5$ dimensions, the event horizon and charge of the black hole are explicitly found in terms of the mass and the (A)dS radius. In any $D (\geq 4)$ dimensions the general relations for black hole radius, mass and charge are used to explore the Schwinger effect. In Sec.~\ref{Con sec}, we summarize the results and discuss the physical implications.

\section{(Near-) Extremal Nonrotating Charged Black Holes} \label{MP sec}

The Einstein-Maxwell theory with a cosmological constant ($\pm$ for AdS/dS) in a $D = n + 3$ dimensional spacetime (in unit of $c = \hbar = 1$)
\begin{eqnarray}
S = \int d^D x \sqrt{-g} \Bigl[ \frac{1}{16 \pi G_D} \Bigl( R \pm \frac{(D - 1)(D - 2)}{L^2} \Bigr) - \frac{1}{4} F_{\mu\nu} F^{\mu\nu} \Bigr],
\end{eqnarray}
has the asymptotically flat ($L \to \infty$) Myers-Perry black hole~\cite{Myers:1986un}
\begin{eqnarray} \label{MP-bh}
ds^2 &=& - f(r) dt^2 + \frac{dr^2}{f(r)} + r^2 d\Omega_{n+1}^2, \qquad f(r) = 1 - \frac{16 \pi G_D M}{(n + 1) A_{n+1}} \frac1{r^n} + \frac{8 \pi G_D Q^2}{n(n+1)} \frac1{r^{2n}},
\nonumber\\
A &=& \frac{Q}{n} \frac1{r^n} \, dt.
\end{eqnarray}
Here, $M$ and $Q$ are the mass and charge of the black hole and $A_{n + 1} = 2 \pi^{(n+2)/2}/\Gamma(\frac{n+2}2)$ is the area of $n + 1$-dimensional unit sphere. The MP black hole has two real positive roots of $f(r) = 0$, $r_+$ for the event horizon and $r_-$ the causal (Cauchy) horizon:
\begin{eqnarray}
r_\pm^n = \frac{8 \pi G_D M}{(n + 1) A_{n+1}} \pm \sqrt{\Bigl( \frac{8 \pi G_D M}{(n + 1) A_{n+1}} \Bigr)^2 -  \frac{8 \pi G_D Q^2}{n(n+1)}}. 
\end{eqnarray}
The Hawking temperature is
\begin{eqnarray}
T_\mathrm{H} = \frac{n \bigl(r_+^{n} - r_-^{n} \bigr)}{4 \pi r_+^{n+1}}.
\end{eqnarray}
In the extremal limit, two horizons degenerate
\begin{eqnarray}
M = M_0 \equiv \sqrt{\frac{n + 1}{8 \pi n G_D}} A_{n+1} Q \quad \to \quad r_+^n = r_-^n = r_0^n \equiv \frac{8 \pi G_D M_0}{(n + 1) A_{n+1}}.
\end{eqnarray}

For the near-extremal black holes
\begin{eqnarray}
r_\pm = r_0 \pm \epsilon B, \qquad M = M_0 + \epsilon^2 B^2 \frac{n^2 M_0}{2 r_0^2},
\end{eqnarray}
the near-horizon geometry is obtained by taking $\epsilon \rightarrow 0$ of the transformation
\begin{eqnarray}
r = r_0 + \epsilon \rho, \qquad t = \frac{\tau}{\epsilon}.
\end{eqnarray}
The extremal black hole corresponds to $B = 0$. Then, the Myers-Perry metric takes the geometry of $\mathrm{AdS}_2 \times \mathrm{S}^{n+1}$
\begin{eqnarray} \label{MP new}
ds^2 = - \frac{\rho^2 - B^2}{R_\mathrm{A}^2} d\tau^2 + \frac{ R_\mathrm{A}^2}{\rho^2 - B^2} d\rho^2 + R_\mathrm{S}^2 d\Omega_{n+1}^2,
\end{eqnarray}
and the gauge potential
\begin{eqnarray}
A = - \frac{Q \rho}{R_\mathrm{S}^{n+1}} \, d \tau = - \sqrt{\frac{n(n+1)}{8 \pi G_D}} \frac{\rho}{R_\mathrm{S}} \, d\tau,
\end{eqnarray}
where
\begin{eqnarray}
R_\mathrm{S} = r_0, \qquad R_\mathrm{A} = r_0/n.
\end{eqnarray}
In the near-horizon geometry the Hawking temperature and chemical potential become
\begin{eqnarray}
T_\mathrm{H} = \frac{B}{2 \pi R_\mathrm{A}^2}, \qquad \Phi_\mathrm{H} = \frac{Q B}{R_\mathrm{S}^{n + 1}}.
\end{eqnarray}

A probe charged scalar with mass $m$ and charge $q$ obeys the Klein-Gordon equation
\begin{eqnarray}
\bigl( \nabla_\mu - i q A_\mu \bigr) \bigl( \nabla^\mu - i q A^\mu \bigr) \Phi - m^2 \Phi = 0.
\end{eqnarray}
Separating the energy and angular momentum conserved by the metric (\ref{MP new})
\begin{eqnarray}
\Phi(\tau, \rho, \Omega) = \mathrm{e}^{-i \omega \tau} H_l(\Omega) R(\rho),
\end{eqnarray}
where $H_l(\Omega)$ is the surface harmonic of degree $l$ on $\mathrm{S}^{n+1}$ with $\nabla_{\perp}^2 H_l (\Omega) = -l (l+n) H_l (\Omega)$ \cite{Bateman:1955}, the radial equation becomes
\begin{eqnarray} \label{rad eq}
\Biggl[ \partial_{\rho} \bigl( \rho^2 - B^2 \bigr) \partial_{\rho} +  R_\mathrm{A}^4 \frac{\bigl( \omega - q E_\mathrm{S} \rho \bigr)^2}{\rho^2 - B^2} - R_\mathrm{A}^2 \Bigl( m^2 + \frac{l (l + n)}{R_\mathrm{S}^2} \Bigr) \Biggr] R(\rho) = 0.
\end{eqnarray}
Here $E_\mathrm{S}$ is the (constant) electric field at $R_\mathrm{S}$:
\begin{eqnarray}
E_\mathrm{S} = \frac{Q}{R_\mathrm{S}^{n+1}}.
\end{eqnarray}
The solutions for the near-extremal black hole are given by the hypergeometric function as
\begin{eqnarray} \label{hyper sol}
R(\rho) &=& c_1 (\rho - B)^{i a_-} (\rho + B)^{i a_+} F\Bigl( \frac12 + i (a_+ + a_- + b), \frac12 + i (a_+ + a_- - b); 1 + i 2 a_-; z \Bigr)
\nonumber\\
&+& c_2 (\rho - B)^{-i a_-} (\rho + B)^{i a_+} F\Bigl( \frac12 + i (a_+ - a_- + b), \frac12 + i (a_+ - a_- - b); 1 - i 2 a_-; z \Bigr),
\end{eqnarray}
where
\begin{eqnarray}
a_\pm = \frac{\omega \pm q E_\mathrm{S} B}{2B} R_\mathrm{A}^2, \qquad b = \sqrt{\bigl(q E_\mathrm{S} R_\mathrm{A}^2 \bigr)^2 - \Bigl( m^2 + \frac{l (l + n)}{R_\mathrm{S}^2} + \frac{1}{4 R_\mathrm{A}^2} \Bigr) R_\mathrm{A}^2}, \qquad z = - \frac{\rho - B}{2B}.
\end{eqnarray}
The extremal  black hole $(B=0)$ has the solutions in terms of the Whittaker functions as
\begin{eqnarray}
R(\rho) = c_1 M\Bigl( i (a_+ - a_-), i b; 2 i \frac{\omega R_\mathrm{A}^2}{\rho} \Bigr) + c_2 W\Bigl( i (a_+ - a_-), i b; 2 i \frac{\omega R_\mathrm{A}^2}{\rho} \Bigr).
\end{eqnarray}

\section{Schwinger Effect in Asymptotically Flat Space} \label{Sch sec}

Following Refs.~\cite{Chen:2012zn, Chen:2014yfa, Chen:2020mqs}, we find the mean number for pair production
\begin{eqnarray} \label{near-sch pair}
\mathcal{N} = \frac{\sinh(2 \pi \mu) \sinh(\pi \tilde{\kappa} - \pi \kappa)}{\cosh(\pi \kappa + \pi \mu) \cosh(\pi \tilde{\kappa} - \pi \mu)} = \frac{\mathrm{e}^{-2 \pi (\kappa - \mu)} - \mathrm{e}^{-2 \pi (\kappa + \mu)}}{1 + \mathrm{e}^{-2 \pi (\kappa + \mu)}} \times  \frac{1 - \mathrm{e}^{-2 \pi (\tilde{\kappa} - \kappa)}}{1 + \mathrm{e}^{-2 \pi (\tilde{\kappa} - \mu)}},
\end{eqnarray}
where
\begin{eqnarray}
\kappa = a_+ - a_- = q E_{\rm S} R_{\rm A}^2 = \frac{q \Phi_{\rm H}}{2 \pi T_{\rm H}}, \qquad \tilde{\kappa} = a_+ + a_- = \frac{\omega R_{\rm A}^2}{B} = \frac{\omega}{2 \pi T_{\rm H}}
\end{eqnarray}
and
\begin{eqnarray} \label{mu2}
\mu^2 = b^2 = \kappa^2 - \bar{m}^2 R_{\rm A}^2, \qquad \bar{m}^2 =  m^2 + \frac{(l + n/2)^2}{R_\mathrm{S}^2}.
\end{eqnarray}
Note that $\bar{m}$ plays the role of an effective mass; the s-wave ($l = 0$) in $D=4$ dimensions, for instance, has $\bar{m}$ in the $\mathrm{AdS}_2$ space~\cite{Cai:2014qba}. Then, the mean number has a universal formula for a thermal interpretation~\cite{Kim:2015kna}
\begin{eqnarray} \label{univ for}
\mathcal{N} = \underbrace{\Biggl( \frac{\mathrm{e}^{- \bar m/T_\mathrm{S}} - \mathrm{e}^{- \bar m/\bar T_\mathrm{S}}}{1 + \mathrm{e}^{- \bar m/\bar T_\mathrm{S}}} \Biggr)}_\textrm{Schwinger effect in AdS$_2$} \times \Biggl\{ \mathrm{e}^{\bar m/T_\mathrm{S}} \underbrace{\Biggl( \mathrm{e}^{- \bar m/T_\mathrm{S}} \frac{1 - \mathrm{e}^{-(\omega - q \Phi_\mathrm{H})/T_\mathrm{H}}}{1+  \mathrm{e}^{- \bar m/T_\mathrm{S}} \mathrm{e}^{-(\omega - q \Phi_\mathrm{H})/T_\mathrm{H}}} \Biggr)}_\textrm{Schwinger effect in Rindler$_2$} \Biggr\},
\end{eqnarray}
where
\begin{eqnarray} \label{eff tem}
T_\mathrm{S} &=& \frac{\bar m}{2 \pi (\kappa - \mu)} = T_\mathrm{U} + \sqrt{T_\mathrm{U}^2 - \frac1{4 \pi^2 R_\mathrm{A}^2}}, \qquad T_\mathrm{U} = \frac{q E_\mathrm{S}}{2 \pi \bar{m}},
\nonumber\\
\bar{T}_\mathrm{S} &=& \frac{\bar m}{2 \pi (\kappa + \mu)} = T_\mathrm{U} - \sqrt{T_\mathrm{U}^2 - \frac1{4 \pi^2 R_\mathrm{A}^2}}.
\end{eqnarray}
Here, $T_\mathrm{U}$ has the meaning of the Unruh temperature for accelerated charge on the spherical surface $R_\mathrm{S}$ and the square root is the Unruh temperature in the $\mathrm{AdS}_2$ space~\cite{Deser:1997ri}.

A few comments are in order. First, the first parenthesis is the Schwinger effect for an extremal black hole, which is the Schwinger effect in the $\mathrm{AdS}_2$ space~\cite{Cai:2014qba}, while the second curly bracket is the additional Schwinger for a near-extremal black hole. The universal form (\ref{univ for}) has been shown for RN black holes~\cite{ Chen:2012zn, Chen:2014yfa}, KN black holes~\cite{Chen:2016caa, Chen:2017mnm}, and RN-(A)dS black holes~\cite{Chen:2020mqs} in four dimensions. Second, the Hawking radiation and the Schwinger effect are intertwined for near-extremal black holes. The surface gravity of the event horizon gives the acceleration of the two-dimensional Rindler space, and the effective temperature $T_\mathrm{S}$ and the Hawking temperature $T_\mathrm{H}$ determine the QED effect in the electric field of charged black hole. The pair production in extremal RN black holes is the limit of $B = 0$ and thereby $T_\mathrm{H} = 0$ and the second factor of curly bracket becomes unity. Third, pairs are produced when the Breitenlohner-Freedman (BF) bound~\cite{Breitenlohner:1982jf, Breitenlohner:1982bm} is violated, $T_\mathrm{U} \geq 1/2\pi R_\mathrm{A}$, which leads to
\begin{eqnarray} \label{BF boun}
q \geq \sqrt{\frac{8 \pi n G_D }{n+1}} \bar{m}.
\end{eqnarray}
A passing remark is that the mean number (\ref{near-sch pair}) is related to the absorption cross section ratio of the field $R(\rho)$ via $\mathcal{N}=-\sigma_{\rm abs}(\mu\rightarrow -\mu)$~\cite{Chen:2016caa,Chen:2017mnm,Zhang:2020apg} and the CFT scalar operator with a complex conformal weight dual to $R(\rho)$ gives the holographic description of RN black holes in arbitrary $D(\geq 4)$-dimensional asymptotically flat or (A)dS space.

\begin{figure}[htbp!]
\centering
\includegraphics[width = 0.6\textwidth]{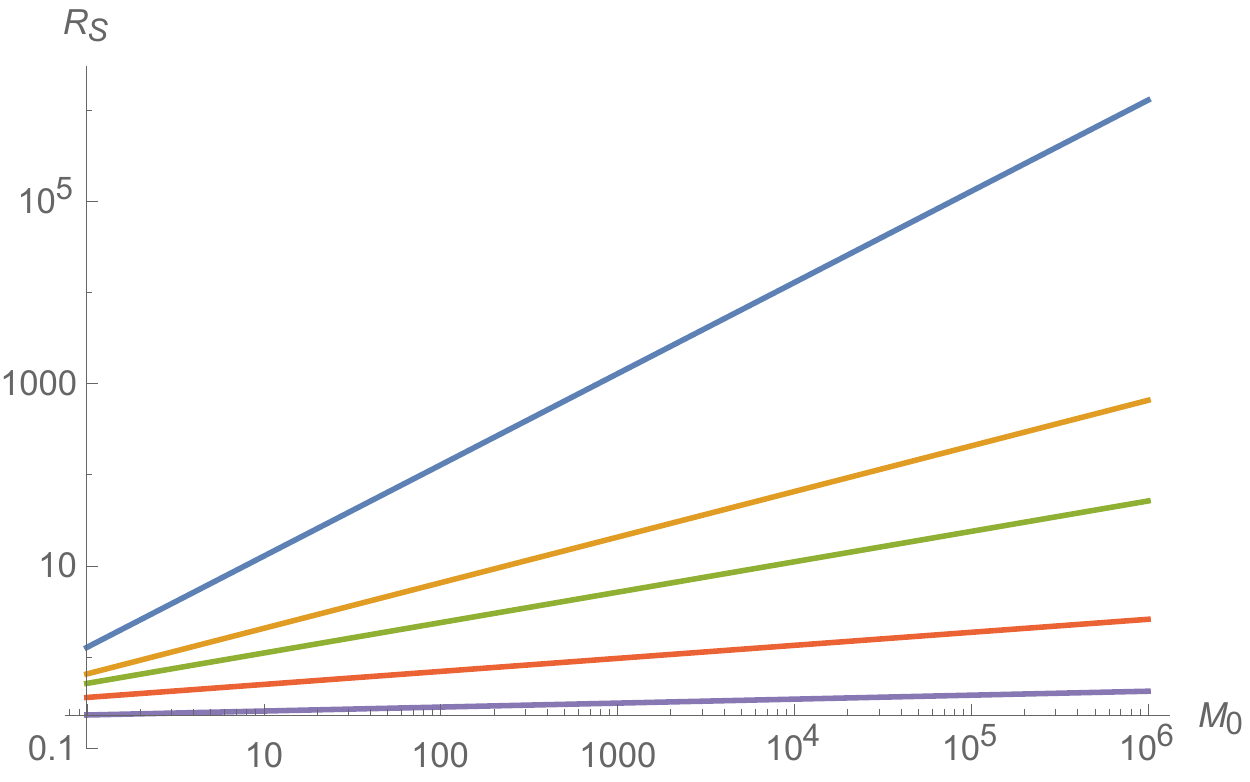}
\caption{\label{R-flat}
The radius $R_\mathrm{S}(M_0, n)$ in the range of $[1, 10^6]$ of $M_0$ for various dimensions (unit of $G_D = 1$): $D = 4$ (blue), $D = 5$ (yellow), $D = 6$ (green), $D = 10$ (red), and $D = 26$ (purple).}
\end{figure}

\begin{figure}[h]
\includegraphics[width=0.45\linewidth,height=0.35\textwidth]{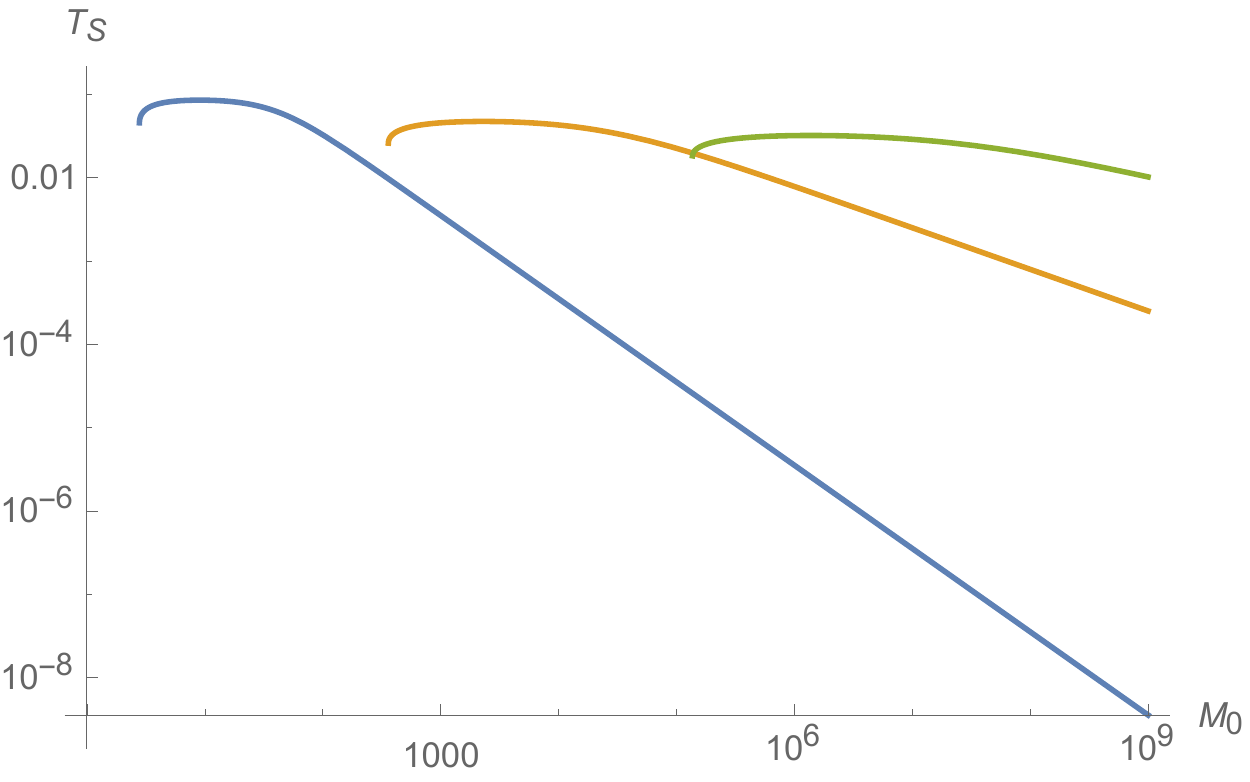}
\hfill
\includegraphics[width=0.45\linewidth,height=0.35\textwidth]{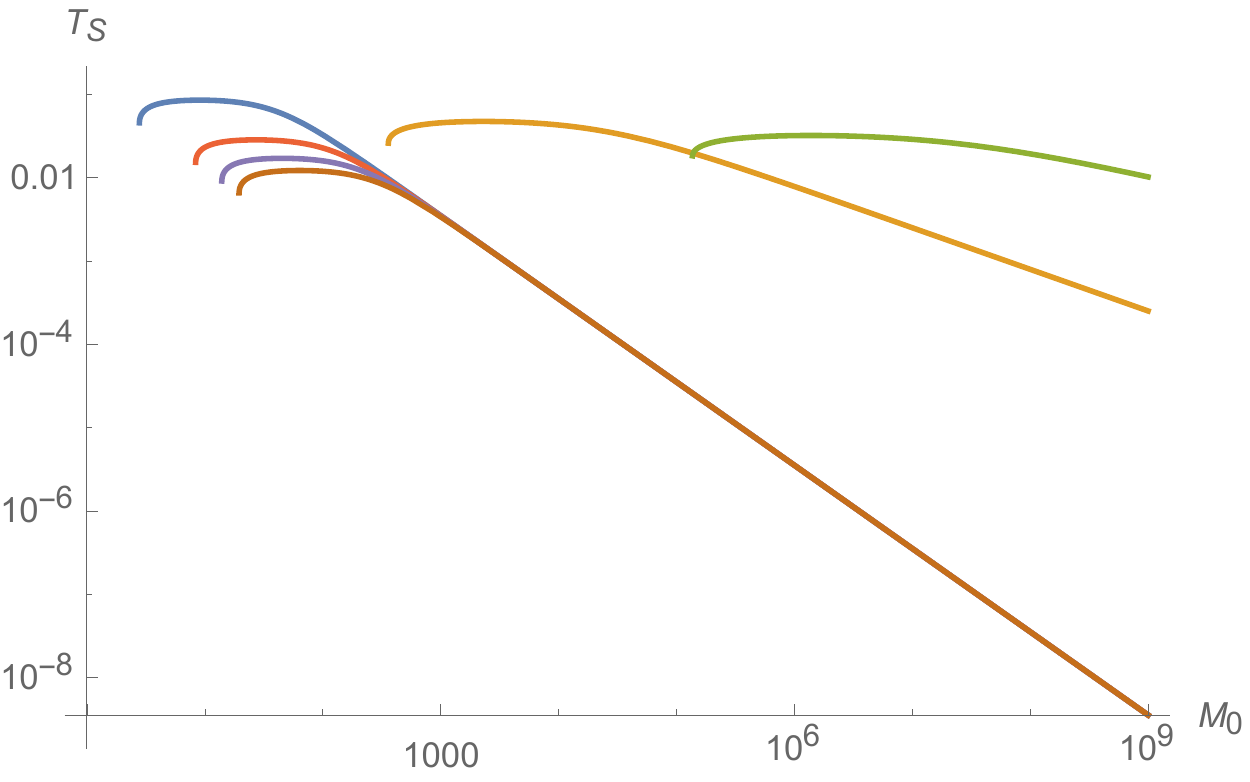}
\caption{\label{T-flat}
[Left panel] The effective temperature $T_\mathrm{S}$ in the range of $[1, 10^9]$ of $M_0$ for an accelerated charge with charge $q = 1$, mass $m = 0.01$ and angular momentum $l = 0$ for dimensions $D = 4$ (blue), $D = 5$ (yellow), and $D = 6$ (green). [Right panel] The $T_\mathrm{S}$'s in $D = 4$ with $l = 0$ (blue), $D = 5$ with $l = 0$ (yellow), $D = 6$ with $l = 0$ (green) are compared with those in $D = 4$ with angular momentum $l = 1$ (red), $D = 4$ with $l = 2$ (purple), $D = 4$ with $l = 3$ (brown).}
\end{figure}

Figures~\ref{R-flat} and \ref{T-flat} show the dimensionality of the horizon radius $R_\mathrm{S}$ and the effective temperature $T_\mathrm{S}$ in Eq.~(\ref{eff tem}). As shown in Fig.~\ref{R-flat}, all radii $R_\mathrm{S}$ increase as $M_0$ and for large $D$ dimensions, $R_\mathrm{S}$ increases as a power-law for large $M_0$. The larger the $D$ dimensions are, the smaller the radius $R_\mathrm{S}$ is. The Unruh temperature in Eq.~(\ref{eff tem}) can be written as
\begin{eqnarray} \label{TU}
T_\mathrm{U} = \sqrt{\frac{n(n+1)}{32 \pi^3 G_D}} \times \frac{q}{\sqrt{(mR_\mathrm{S})^2 + \bigl(l+ n/2 \bigr)^2}}.
\end{eqnarray}
For a given large $M_0$ the effective temperature $T_\mathrm{S}$ for high dimensions is higher than that for low dimensions since the radius $R_\mathrm{S}$ for the charge acceleration by electric field rapidly increase for $D=4$ dimensions while it slowly increases for $D \geq 5$ dimensions, which is illustrated in Fig.~\ref{T-flat}. Note that the BF bound, below which the effective temperature is not defined and pairs are not produced, increases as $M_0$ by order of magnitude for high dimensions as shown in the left panel of Fig.~\ref{T-flat}. The $T_\mathrm{S}$ has the maximum, which depends on the dimensionality. Larger angular momenta suppress $T_\mathrm{S}$ for small mass black holes but they approaches the same temperature for large mass black holes, as shown in the right panel of Fig.~\ref{T-flat}. The leading Boltzmann factor of pair production is ${\cal N} \approx \mathrm{e}^{-\bar{m}/T_\mathrm{S}}$ and the exponent is $0.01/T_\mathrm{S}$ in Fig.~\ref{T-flat}. This implies that in $D=4$ dimensions the Schwinger effect is the most efficient for small (Planck scale) (near-) extremal RN black holes while it is exponentially suppressed for large black holes. On the other hand, the BF bound is order of $10^3$ for $D=5$ dimensions and $10^5$ for $D=6$ dimensions, beyond which the Schwinger effect is profound and produces a cornucopia of pairs near the event horizon.
Note that any black hole within the BF bound is stable against Schwinger pair production, not to mention the Hawking radiation, and that the BF bound increases for high dimensions by order of magnitude.

\section{(Anti-) de Sitter Space} \label{CC sec}

For the asymptotical (A)dS black holes, the function $f(r)$ in the metric has the form~\cite{Xu:1988ju}
\begin{eqnarray}
f(r) = 1 - \frac{16 \pi G_D M}{(n + 1) A_{n+1}} \frac1{r^n} + \frac{8 \pi G_D Q^2}{n(n+1)} \frac1{r^{2n}} \pm \frac{r^2}{L^2}.
\end{eqnarray}
The extremal condition requires a constraint on mass and charge implicitly via the radius of the degenerate horizon $r_0$ by coinciding the event and causal horizons
\begin{eqnarray} \label{AdS m-q}
M_0 = \frac{(n+1) A_{n+1}}{8 \pi G_D} r_0^n \left( 1 \pm \frac{n + 1}{n} \frac{r_0^2}{L^2} \right), \qquad Q^2 = \frac{n (n+1)}{8 \pi G_D} r_0^{2n} \left( 1 \pm \frac{n + 2}{n} \frac{r_0^2}{L^2} \right).
\end{eqnarray}
The other extremal black holes in the dS space are Nariai black holes, which are obtained by coinciding the event and cosmological horizons~ \cite{Cardoso:2004uz}.
In $D = 4$ dimensions, the degenerate event horizon and mass can be explicitly expressed in terms of charge $Q$ and the radius $L$ as discussed in Ref.~\cite{Chen:2020mqs}. Similarly, the event and cosmological horizons of extremal $\mathrm{RN-(A)dS}_5$ are explicitly found in terms of mass $M_0$
\begin{eqnarray} \label{d=5 hor}
r_0^2 = \frac{L^2}3 \delta, \qquad r_\mathrm{C}^2 = \mp L^2 \Bigl( 1 \pm \frac23 \delta \Bigr), \qquad \delta = \pm \Bigl( \sqrt{1 \pm \frac{8 G_5 M_0}{\pi L^2}} - 1 \Bigr),
\end{eqnarray}
and the charge is
\begin{eqnarray}
Q^2 = \frac{L^4}{36 \pi G_5} \delta^2 (3 \pm 2 \delta),
\end{eqnarray}
where the upper (lower) sign is for the AdS (dS) space. Note that the cosmological horizon does not exist in the AdS space since $r_\mathrm{C}^2 < 0$, as expected.

In any $D(\geq 4)$ dimensions for the mass slightly larger than $M_0$
\begin{eqnarray}
M = M_0 + \epsilon^2 B^2 \frac{(n+1) A_{n+1}}{16 \pi G_D} \frac{R_\mathrm{S}^n}{R_\mathrm{A}^2},
\end{eqnarray}
we have the near horizon geometry~(\ref{MP new}) with
\begin{eqnarray}
R_\mathrm{S} = r_0, \qquad R_\mathrm{A} = \frac{R_\mathrm{S}}{\sqrt{n^2 \pm (n+1)(n+2) R_\mathrm{S}^2/L^2}}.
\end{eqnarray}
The results for asymptotically flat case apply to the (A)dS space by using modified ratio of $R_\mathrm{A}/R_\mathrm{S}$, namely replacing  $(l + n/2)^2 \to (l + n/2)^2 \pm (n+1) (n+2) R_\mathrm{S}^2/L^2$ in Eqs.~(\ref{mu2}), (\ref{BF boun}) and (\ref{TU}):
\begin{eqnarray} \label{BF-AdS}
q \geq \sqrt{\frac{8 \pi n G_D }{n+1}} \bar{m} \times \sqrt{\frac{1 \pm \frac{(n+1)(n+2)}{n^2} \frac{R_\mathrm{S}^2}{L^2}}{1 \pm \frac{n+2}{n} \frac{R_\mathrm{S}^2}{L^2}}}, \qquad T_\mathrm{U} = \sqrt{\frac{n(n+1)}{32 \pi^3 G_D}} \times \frac{q}{\bar{m} R_\mathrm{S}} \times \sqrt{1 \pm \frac{n+2}{n} \frac{R_\mathrm{S}^2}{L^2}},
\end{eqnarray}
where
\begin{eqnarray}
\bar{m}^2 = m^2 + \frac{(l + n/2)^2}{R_\mathrm{S}^2} \pm \frac{(n+1)(n+2)}{4 L^2}.
\end{eqnarray}
The BF bounds are discussed in the $\mathrm{AdS}_2$ space~\cite{Pioline:2005pf} and in the $\mathrm{(A)dS}_2$ space~\cite{Kim:2008xv,Cai:2014qba}.


\begin{figure}[h]
\includegraphics[width=0.45\linewidth,height=0.35\textwidth]{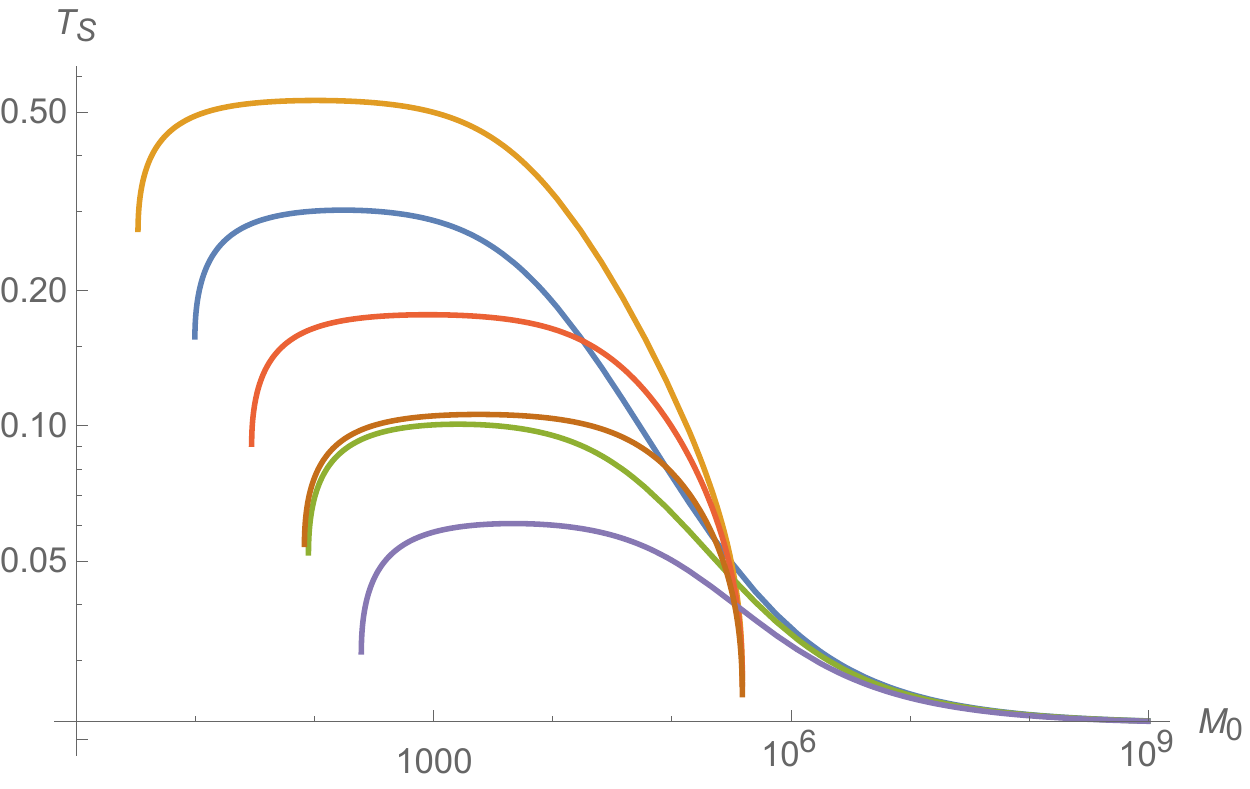}
\hfill
\includegraphics[width=0.45\linewidth,height=0.35\textwidth]{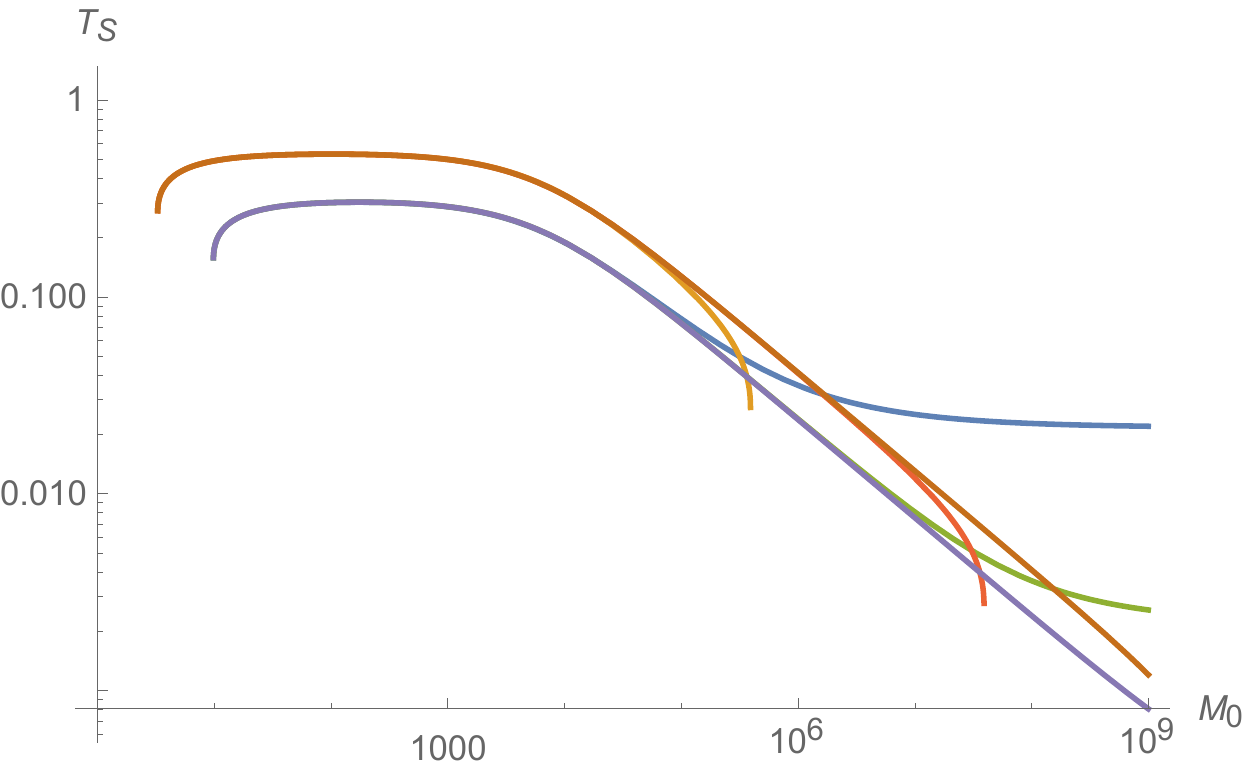}
\caption{\label{T-S-M-C}
[Left panel] The $T_\mathrm{S}$ in $D=5$ (A)dS space in the range of $[1, 10^9]$ of $M_0$ for a charge with $q = 1$, $m = 0.01$ and the (A)dS radius $L = 10^3$ for AdS space with $l=0$ (blue), dS space with $l = 0$ (yellow), AdS space with $l = 1$ (green), dS space with $l=1$ (red), AdS space with $l = 2$ (purple), and dS space with $l=2$ (brown). [Right panel] The $T_\mathrm{S}$ in $D=5$ (A)dS space for a charge in $l=0$ for AdS space with $L = 10^3$ (blue), dS space with $L=10^3$ (yellow), AdS space with $L=10^4$ (green), dS space with $L=10^4$ (red), AdS space with $L=10^5$ (purple), and dS space with $L=10^5$ (brown).}
\end{figure}

For dimensions higher than $D=5$, the explicit expression for $R_\mathrm{S}$ cannot be found. However, using the implicit relation (\ref{AdS m-q}), we can draw figures of the effective temperature $T_\mathrm{S}$ for an accelerated charge as functions of $R_\mathrm{S}$ for various dimensions. The event horizon is explicitly found as a function of $M_0$ for $D=5$ dimensions in Eq.~(\ref{d=5 hor}) while one can implicitly find a general relation (\ref{AdS m-q}) for any $D (\geq 4)$ dimensions. In Fig.~\ref{T-S-M-C} we numerically calculate the effective temperature, $T_\mathrm{S}$, for (near-) extremal RN black holes in $D=5$ dimensions for the asymptotically (A)dS space and compare them between the $\mathrm{dS}_5$ and $\mathrm{AdS}_5$ spaces.

First, the (near-) extremal five-dimensional RN black holes in the asymptotically dS space have the lower bound (\ref{BF-AdS}) and another bound from the existence of event horizon for pair production
\begin{eqnarray}\label{dS boun}
L \geq \sqrt{\frac{8 G_5 M_0}{\pi}}.
\end{eqnarray}
For numerical purpose, we assume that the cosmological horizon is far beyond the event horizon and that the Schwinger effect from the cosmological horizon is negligible. As shown in Fig.~\ref{T-S-M-C}, the effective temperature in the $\mathrm{(A)dS}_5$ space rapidly increases beginning from the BF bound, reaches an almost plateau region for a wide range of $M_0$ and then decreases. The effective temperature for dS space rapidly decreases near the dS bound (\ref{dS boun}) while that for AdS space decreases slowly for large $M_0$.  For a given mass parameter $M_0$ and the (A)dS radius $L$, the effective temperature for the $\mathrm{dS}_5$ space is higher than that for the $\mathrm{AdS}_5$ space until the $M_0$ reaches the dS bound (\ref{dS boun}), and the effective temperature and the Schwinger pair production for a charge in $l(\geq 1)$ states are suppressed for both AdS and dS space, as shown in the left panel of Fig.~\ref{T-S-M-C}. And the right panel of Fig.~\ref{T-S-M-C} shows that the difference of the effect of the (A)dS radius is negligible for small $M_0$ but the larger AdS radius $L$ suppresses the effective temperature for large $M_0$.

\begin{figure}[h]
\includegraphics[width=0.45\linewidth,height=0.35\textwidth]{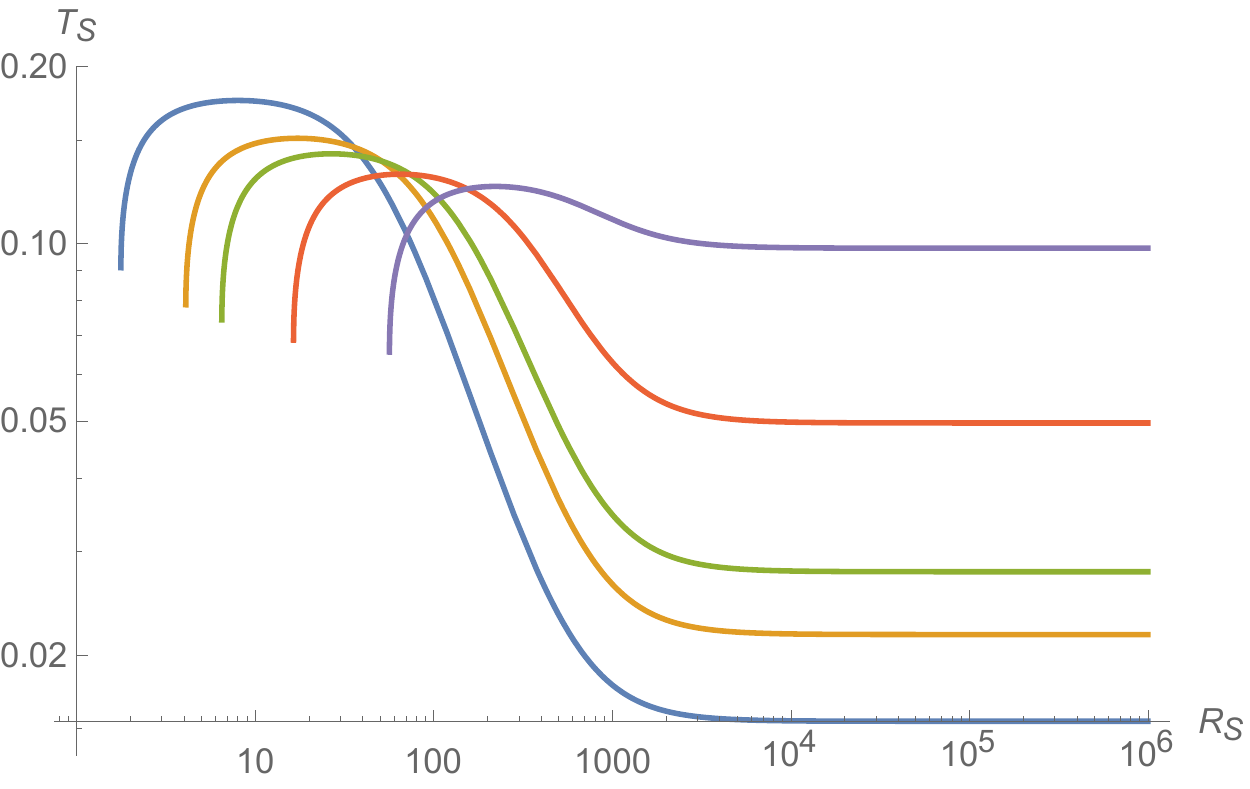}
\hfill
\includegraphics[width=0.45\linewidth,height=0.35\textwidth]{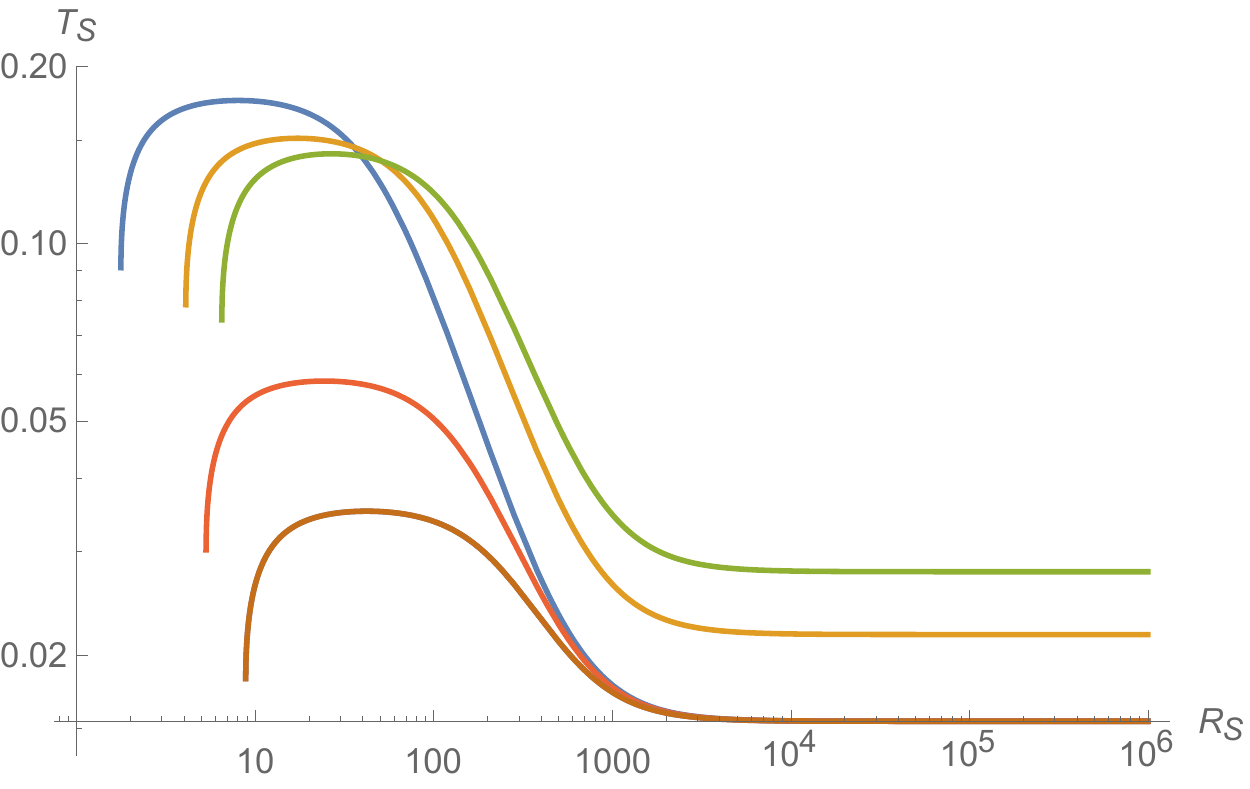}
\caption{\label{T-S-R-S-AdS-C}
[Left panel] The $T_\mathrm{S}$ in AdS space in the range of $[1, 10^6]$ of $R_\mathrm{S}$ for a charge with $q = 1$, $m = 0.01$, $l=0$ and the AdS radius $L = 10^3$ for dimensions $D = 4$ (blue), $D = 5$ (yellow), $D = 6$ (green), $D = 10$ and $D = 26$. [Right panel] The $T_\mathrm{S}$ in AdS space for dimensions $D = 4$ with $l = 0$ (blue), $D = 5$ with $l = 0$ (yellow), and $D = 6$ with $l = 0$ (green), and $D = 4$ with $l = 1$ (red), $D = 4$ with $l = 2$ (purple), $D = 4$ with $l = 3$ (brown).}
\end{figure}

Second, the (near-) extremal RN black holes in the asymptotically AdS space have the BF bound (\ref{BF-AdS}) but does not have a cosmological horizon and the dS bound (\ref{dS boun}), as shown in Fig.~\ref{T-S-R-S-AdS-C}, in which the AdS radius is fixed to $L= 10^3$. Figure~\ref{T-S-R-S-AdS-C} is the effective temperature, $T_\mathrm{S}$, for $R_\mathrm{S}$ up to $10^6$ in unit of $G_D =1$.
The peak values of $T_\mathrm{S}$'s for low dimensions are slightly higher than those for high dimensions but $T_\mathrm{S}$'s for low dimensions decrease after reaching their peaks and cross those for higher dimensions. Interestingly, all $T_\mathrm{S}$'s approach plateaus for large $R_\mathrm{S}$. As shown in the right panel, the quantum states for charge with $l \geq 1$ are more suppressed than the s-wave and $l$ affects more than the dimensionality. This implies that even very large (near-) extremal RN black holes in the AdS space efficiently produce pairs dominantly in $s$-wave and become instantaneously unstable due to the Schwinger effect.

\begin{figure}[h]
\includegraphics[width=0.45\linewidth,height=0.35\textwidth]{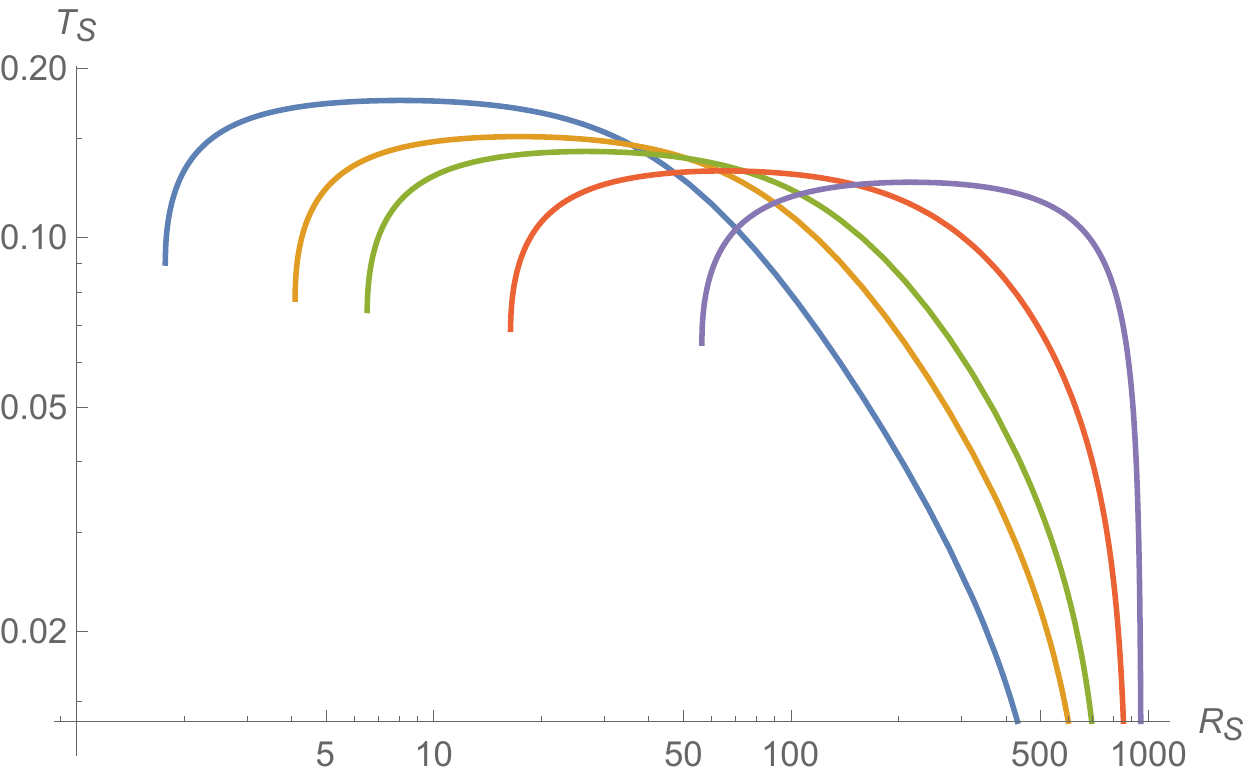}
\hfill
\includegraphics[width=0.45\linewidth,height=0.35\textwidth]{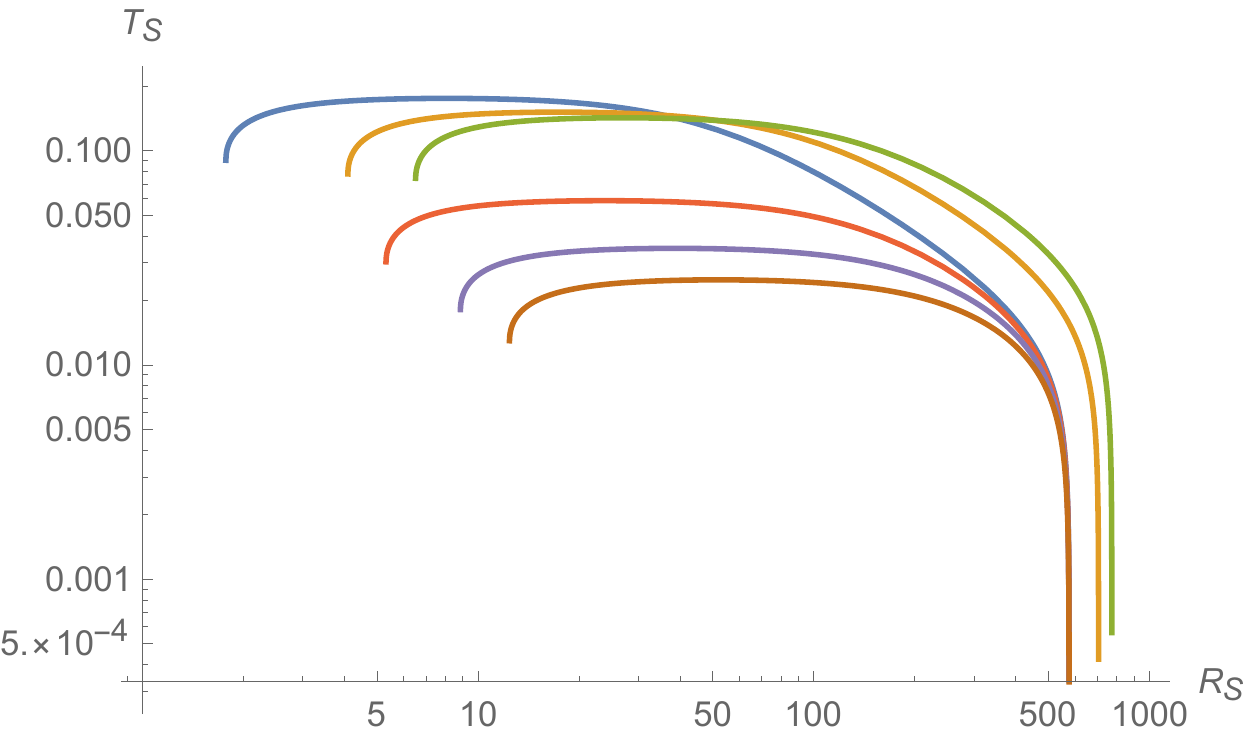}
\caption{\label{T-S-R-S-dS-C}
[Left panel] The $T_\mathrm{S}$ in dS space in the range of $[1, 10^6]$ of $R_\mathrm{S}$ for a charge with $q = 1$, $m = 0.01$, $l = 0$ and the dS radius $L = 10^3$ for dimensions $D = 4$ (blue), $D = 5$ (yellow), $D = 6$ (green), $D=10$ (red) and $D=26$ (purple). [Right panel] The $T_\mathrm{S}$ in dS space for dimensions $D = 4$ with $l=0$ (blue), $D = 5$ with $l=0$ (yellow), $D = 6$ with $l=0$ (green), $D=4$ with $l=1$ (red), $D=4$ with $l=2$ (purple) and $D=4$ with $l=3$ (brown).}
\end{figure}

Third, Fig.~\ref{T-S-R-S-dS-C} shows the effective temperature $T_\mathrm{S}$ for the dS space. The dS space has two bounds for the Schwinger effect: the BF bound and the dS bound. The RN-dS black holes can emit charged particles only within these two bounds. The BF bound gives small black holes an upper bound for the stability against the Schwinger effect while the dS bound gives large black holes a lower bound for the stability. The dS bound strongly contrasts to the no-bound for large black holes in the asymptotically flat or AdS space. The higher the spacetime dimensions are, the larger the BF and dS bounds are. The Schwinger effect is also an efficient mechanism for a wide range of black holes in all dimensional spacetime since $T_\mathrm{S}$ keeps almost the same order up to the dS bound. The emission of charges in high angular momentum is exponentially suppressed.

\begin{figure}[h]
\includegraphics[width=0.45\linewidth,height=0.35\textwidth]{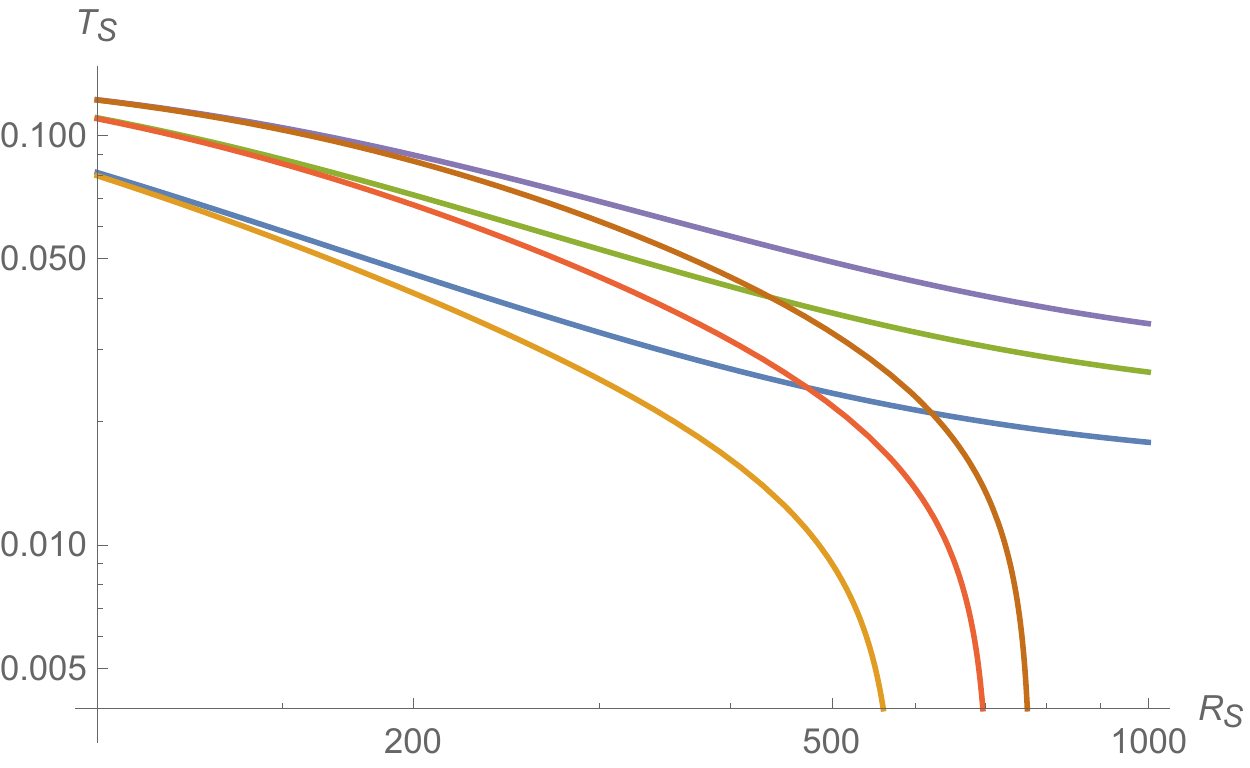}
\hfill
\includegraphics[width=0.45\linewidth,height=0.35\textwidth]{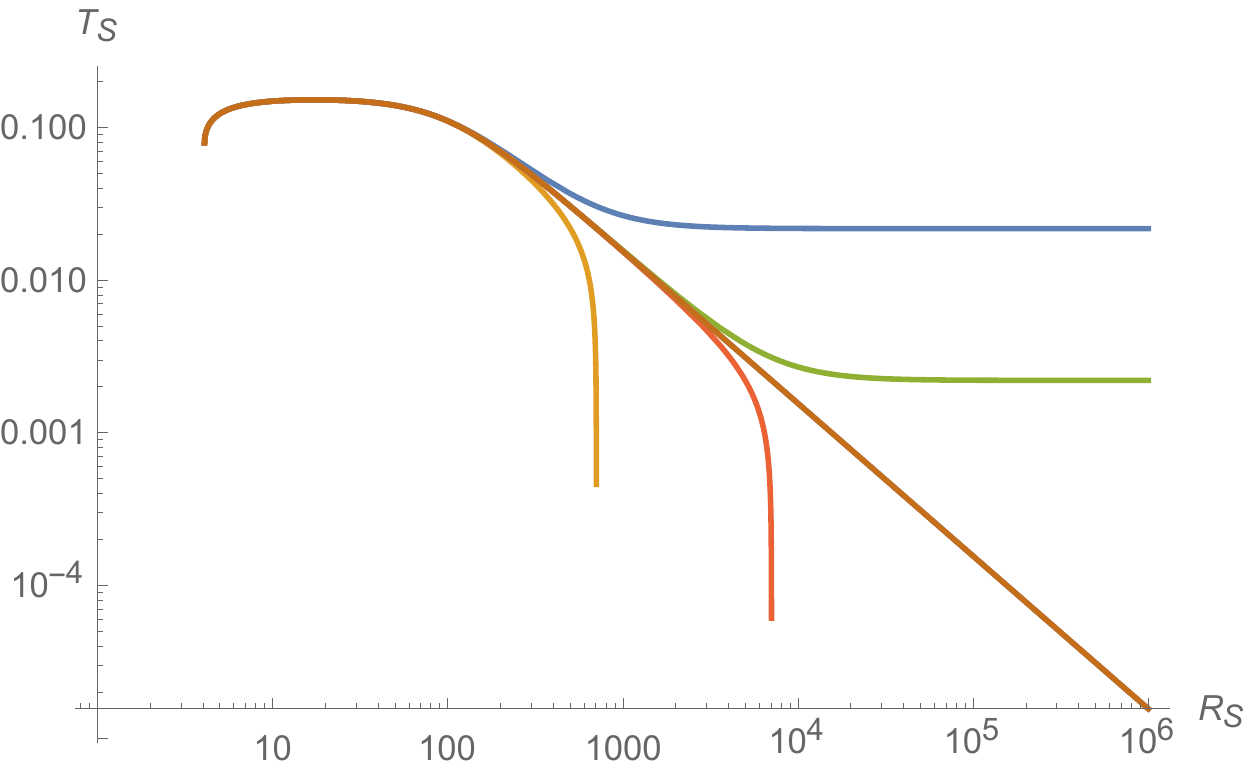}
\caption{\label{T-S-R-S-C}
[Left panel] The $T_\mathrm{S}$ in (A)dS space in the range of $[1, 10^3]$ of $R_\mathrm{S}$ for a charge with $q = 1$, $m = 0.01$, $l = 0$ and the (A)dS radius $L = 10^3$ for $D = 4$ AdS space (blue), $D = 4$ dS space (yellow), $D = 5$ AdS space (green), $D=5$ dS space (red), $D=6$ AdS space (purple) and $D=6$ dS space (brown). [Right panel] The $T_\mathrm{S}$ in $D=5$ in the range of $[1, 10^6]$ for AdS space with $L=10^3$ (blue), dS space with $L=10^3$ (yellow), AdS space with $L = 10^4$ (green), dS space with $L=10^4$ (red) and AdS space with $L = 10^8$ (purple), dS space with $L=10^8$ (purple).}
\end{figure}

Finally, we compare the effective temperature $T_\mathrm{S}$'s for a charge as functions of the black hole radius between the AdS and dS spaces. Compared to the $\mathrm{(A)dS}_4$ space in Ref.~\cite{Chen:2020mqs} and $\mathrm{(A)dS}_5$ space in Sec.~\ref{CC sec}, the radius $R_\mathrm{S}$ of black holes can be expressed implicitly through the relation (\ref{AdS m-q}). In fact, the radius $R_\mathrm{S}$ is a function of $M_0$, $Q$ and $L$. Figure~\ref{T-S-R-S-C} compares $T_\mathrm{S}$'s for an accelerated charge on the event horizon for fixed (A)dS radius $L$. It is numerically shown that $T_\mathrm{S}$ for AdS space is higher than that for dS space for given dimensions, which is analytically shown in $\mathrm{(A)dS}_4$ space~\cite{Chen:2020mqs}. For small black holes the Schwinger effect from AdS space does not differ from that from dS space, but the difference gets larger for large black holes, in particular, near the dS bound. The higher dimensional (A)dS space has higher $T_\mathrm{S}$, which implies that the higher dimensional (A)dS space is more unstable against the Schwinger effect than lower dimensional (A)dS space. The (A)dS radius $L$ does not affect $T_\mathrm{S}$ much for small black holes but large $L$ decreases $T_\mathrm{S}$ for large black holes, which has been observed in Ref.~\cite{Chen:2020mqs}. Contrary to a prejudice that the Gibbons-Hawking radiation in dS space may cooperate to enhance the Schwinger effect, the dS radius $L$ in the asymptotic space increases the horizon radius of black holes while the AdS radius $L$ decreases the black hole radius. In fact, as shown in the right panel of Fig.~\ref{T-S-R-S-C} the AdS space has higher effective temperature than the asymptotically flat space, which is numerically realized by taking a very large (A)dS radius and has higher effective temperature than the dS space.

\section{Conclusion}\label{Con sec}

We have studied the Schwinger effect from (near-) extremal RN black holes in asymptotically flat and (A)dS spacetime in high $(D \geq 4)$ dimensions. The near-extremal RN black holes have a vanishingly small Hawking temperature and suppress the emission of charges through the Hawking radiation. It has been argued since the discovery of Hawking radiation that Schwinger pair production of charged pairs operates even for (near-) extremal black holes. The near-horizon geometry of $\mathrm{AdS}_2$ and warped $\mathrm{AdS}_3$ for near-extremal RN and KN black holes, respectively, in four-dimensional asymptotically flat spacetime allows one to find the analytical formulae for the emission of electric and/or magnetic charges~\cite{Chen:2012zn, Chen:2014yfa,Chen:2016caa, Chen:2017mnm}. In this paper, we have extended the Schwinger effect to RN black holes in high dimensions with/without the asymptotic (A)dS boundary, which have been motivated by string theory and extra-dimensional physics.

In the asymptotically flat spacetime, $D$-dimensional (near-) extremal RN black holes have the $\mathrm{AdS}_2 \times S^{D-2}$ geometry in the near-horizon region. Interestingly, those black holes in $D \geq 5$ dimensions have a different $\mathrm{AdS}_2$ radius $R_\mathrm{A}$ from the event horizon $R_\mathrm{S}$, in strong contrast to four-dimensional RN black holes with the same radius $R_\mathrm{A} = R_\mathrm{S}$. Recently it has been observed that the asymptotic (A)dS boundary makes $\mathrm{AdS}_2$ radius smaller (larger) for the AdS (dS) space than that in asymptotically flat spacetime~\cite{Chen:2020mqs}. The enhanced symmetry of $\mathrm{AdS}_2 \times S^{D-2}$ allows us to find solutions of a charged scalar field and thereby the mean number of charged pairs from the horizon. There is a minimum radius of black hole horizon for each dimensions, above which the near-extremal RN black holes emit charges and whose radius increases by order of magnitude in higher dimensions. The BF bound, below the minimum horizon radius for pair production, is the parameter region of charge and mass of black holes to remain stable against the Schwinger effect as well as the Hawking radiation. Except the Planck scale BF bound in $D=4$ dimensions, the BF bound in $D\geq 5$ dimensions is much larger than the Planck scale in the given dimensions. The effect of dimensionality is more significant than the effect of the excited states of the charge.

In the asymptotic (A)dS space, the event horizon of RN black hole are explicitly found in terms of the mass and charge of the black hole in $D=4$ and $D=5$ dimensions. The implicit relation between the mass, charge and the event horizon still characterizes the Schwinger effect in $D \geq 6$ dimensions. The near-extremal RN black holes have the same near-horizon geometry of $\mathrm{AdS}_2 \times S^{D-2}$, whose $\mathrm{AdS}_2$ radius and the horizon radius have complicated dependence on the (A)dS radius as well as the dimensionality. The asymptotic AdS boundary enhances the Schwinger effect of large RN black holes in $D \geq 5$ dimensions and extends the range of black hole radius for efficient emission than asymptotically flat spacetime, while asymptotic dS boundary suppresses the Schwinger effect of large RN black holes than asymptotically flat spacetime. The reason is that the AdS boundary decreases the black hole radius compared to the asymptotic flat spacetime while the dS boundary increases the black hole radius. Thus the AdS boundary strengthens the electric field near the black hole horizon while the dS boundary weakens the electric field. For given dimensions the effective temperature in asymptotic (A)dS and flat spacetime approaches each other for small RN black holes. The BF bound for the AdS space is slightly larger than that for the dS space and those BF bounds do not differ much from the BF bound in the asymptotically flat spacetime. The dS space has the dS bound from a requirement that the near-extremal black hole should be within the cosmological horizon. The considerable effective temperature for the accelerated charge around the black hole horizon warrants the Schwinger effect for a wide range of black hole radius in $D \geq 5$ dimensions.

Finally, the Schwinger effect from near-extremal RN black holes in $D (\geq 4)$-dimensional (A)dS space has many physical implications.
In $D=4$ dimensions, the dS boundary gives the dS bound, beyond which near-extremal RN black holes remain stable against the Schwinger effect as well as the Hawking radiation, in strong contrast to the Planck scale black holes due to the BF bound in the asymptotically flat space. Further, the quantum field theory for charges may be justified for these large black holes. This opens a possibility for near-extremal, large RN black holes in $\mathrm{dS}_4$ as remnants from the inflationary universe. Not discussed in this paper are the (near-) extremal charged black holes as dark matter~\cite{Bai:2019zcd} or primordial black holes~\cite{Carr:2009jm,Liu:2020cds}. In $D (\geq 5)$ dimensions, the Schwinger effect is more efficient than that in $D=4$ dimensions. It would be worth to generalize the Schwinger effect to the cosmological horizon~\cite{Montero:2019ekk}
and Nariai black holes in high dimensions~\cite{Cardoso:2004uz}. The enhanced Schwinger effect in the AdS boundary makes the AdS space unstable in any gauge field of electric-type, such as electric fields in the Maxwell theory and chromo-electric fields in the Yang-Mills theory. These issues will be addressed in future publications.

\acknowledgments
The work of R.G.C. is supported in part by the National Natural Science Foundation of China Grants No.11690021, No.11690022, No.11851302, No.11821505, and No.11947302, in  part by the Strategic Priority Research Program of the Chinese Academy of Sciences Grant No. XDB23030100, No. XDA15020701 and by Key Research Program of Frontier Sciences, CAS.
The work of C.M.C. was supported by the Ministry of Science and Technology of the R.O.C. under the grant MOST 108-2112-M-008-007.
The work of S.P.K. was supported in part by National Research Foundation of Korea (NRF) funded by the Ministry of Education (2019R1I1A3A01063183).
The work of J.R.S. was supported by the NSFC Grant No. 11675272.


\begin{references}


\bibitem{Hawking:1974sw}
  S.~W.~Hawking,
  ``Particle Creation by Black Holes,''
  Commun.\ Math.\ Phys.\  {\bf 43}, 199 (1975)
  Erratum: [{\it Commun.\ Math.\ Phys.}\  {\bf 46}, 206 (1976)].

\bibitem{Schwinger:1951nm}
  J.~S.~Schwinger,
  ``On gauge invariance and vacuum polarization,''
  Phys.\ Rev.\  {\bf 82}, 664 (1951).

\bibitem{Chen:2012zn}
  C.-M.~Chen, S.~P.~Kim, I.-C.~Lin, J.-R.~Sun and M.-F.~Wu,
  ``Spontaneous Pair Production in Reissner-Nordstrom Black Holes,''
  Phys.\ Rev.\ D {\bf 85}, 124041 (2012)
  [arXiv:1202.3224 [hep-th]].

\bibitem{Chen:2014yfa}
  C.-M.~Chen, J.-R.~Sun, F.-Y.~Tang and P.-Y.~Tsai,
  ``Spinor particle creation in near extremal Reissner–Nordström black holes,''
  Class.\ Quant.\ Grav.\  {\bf 32}, 195003 (2015)
  [arXiv:1412.6876 [hep-th]].

\bibitem{Chen:2016caa}
  C.-M.~Chen, S.~P.~Kim, J.-R.~Sun and F.-Y.~Tang,
  ``Pair Production in Near Extremal Kerr-Newman Black Holes,''
  Phys.\ Rev.\ D {\bf 95}, 044043 (2017)
  [arXiv:1607.02610 [hep-th]].

\bibitem{Chen:2017mnm}
  C.-M.~Chen, S.~P.~Kim, J.-R.~Sun and F.-Y.~Tang,
  ``Pair production of scalar dyons in Kerr–Newman black holes,''
  Phys.\ Lett.\ B {\bf 781}, 129 (2018)
  [arXiv:1705.10629 [hep-th]].

\bibitem{Cai:2014qba}
  R.-G.~Cai and S.~P.~Kim,
  ``One-Loop Effective Action and Schwinger Effect in (Anti-) de Sitter Space,''
  JHEP {\bf 1409}, 072 (2014)
  [arXiv:1407.4569 [hep-th]].

\bibitem{Kim:2019joy}
  S.~P.~Kim,
  ``Astrophysics in Strong Electromagnetic Fields and Laboratory Astrophysics,''
  arXiv:1905.13439 [gr-qc].

\bibitem{Myers:1986un}
  R.~C.~Myers and M.~J.~Perry,
  ``Black Holes in Higher Dimensional Space-Times,''
  Annals Phys.\  {\bf 172}, 304 (1986).

\bibitem{Chen:2020mqs}
  C.-M.~Chen and S.~P.~Kim,
  ``Schwinger Effect from Near-extremal Black Holes in (A)dS Space,''
  arXiv:2002.00394 [hep-th].

\bibitem{Zhang:2020apg}
  J.~Zhang, Y.~Y.~Lin, H.~C.~Liang, K.~J.~Chi, C.~M.~Chen, S.~P.~Kim and J.~R.~Sun,
  ``Pair production from Reissner-Nordstr\"om-anti-de Sitter black holes,''
  arXiv:2003.06398 [hep-th].

\bibitem{Bateman:1955}
  Bateman Project,
  {\it Higher Transcendental Functions}
  (McGraw-Hill Book Company, 1955), Vol. II.

\bibitem{Kim:2015kna}
  S.~P.~Kim, H.~K.~Lee and Y.~Yoon,
  ``Thermal Interpretation of Schwinger Effect in Near-Extremal RN Black Hole,''
  Int.\ J.\ Mod.\ Phys.\ D {\bf 28}, 1950139 (2019)
  [arXiv:1503.00218 [hep-th]].

\bibitem{Deser:1997ri}
  S.~Deser and O.~Levin,
  ``Accelerated detectors and temperature in (anti)-de Sitter spaces,''
  Class.\ Quant.\ Grav.\  {\bf 14}, L163 (1997)
  [gr-qc/9706018].

\bibitem{Breitenlohner:1982jf}
  P.~Breitenlohner and D.~Y.~Freedman,
  ``Stability in Gauged Extended Supergravity,''
  Annals Phys.\  {\bf 144}, 249 (1982).

\bibitem{Breitenlohner:1982bm}
  P.~Breitenlohner and D.~Y.~Freedman,
  ``Positive Energy in anti-De Sitter Backgrounds and Gauged Extended Supergravity,''
  Phys.\ Lett.\  B {\bf 115}, 197 (1982).

\bibitem{Xu:1988ju}
  D.~Y.~Xu,
  ``Exact Solutions of Einstein and Einstein-maxwell Equations in Higher Dimensional Space-time,''
  Class.\ Quant.\ Grav.\  {\bf 5}, 871 (1988).

\bibitem{Cardoso:2004uz}
  V.~Cardoso, O.~J.~C.~Dias and J.~P.~S.~Lemos,
  ``Nariai, Bertotti-Robinson and anti-Nariai solutions in higher dimensions,''
  Phys.\ Rev.\ D {\bf 70}, 024002 (2004)
  [hep-th/0401192].

\bibitem{Pioline:2005pf}
  B.~Pioline and J.~Troost,
  ``Schwinger pair production in AdS(2),''
  JHEP {\bf 0503}, 043 (2005)
  [hep-th/0501169].

\bibitem{Kim:2008xv}
  S.~P.~Kim and D.~N.~Page,
  ``Schwinger Pair Production in dS(2) and AdS(2),''
  Phys.\ Rev.\  D {\bf 78}, 103517 (2008)
  [arXiv:0803.2555 [hep-th]].

\bibitem{Bai:2019zcd}
  Y.~Bai and N.~Orlofsky,
  ``Primordial Extremal Black Holes as Dark Matter,''
  Phys.\ Rev.\ D {\bf 101}, 055006 (2020)
  [arXiv:1906.04858 [hep-ph]].

\bibitem{Carr:2009jm}
  B.~J.~Carr, K.~Kohri, Y.~Sendouda and J.~Yokoyama,
  ``New cosmological constraints on primordial black holes,''
  Phys.\ Rev.\ D {\bf 81}, 104019 (2010)
  [arXiv:0912.5297 [astro-ph.CO]].

\bibitem{Liu:2020cds}
  L.~Liu, Z.-K.~Guo, R.-G.~Cai and S.~P.~Kim,
  ``Merger rate distribution of primordial black hole binaries with electric charges,''
  arXiv:2001.02984 [astro-ph.CO].

\bibitem{Montero:2019ekk}
  M.~Montero, T.~Van Riet and G.~Venken,
  ``Festina Lente: EFT Constraints from Charged Black Hole Evaporation in de Sitter,''
  JHEP {\bf 2001}, 039 (2020)
  [arXiv:1910.01648 [hep-th]].

\end{references}
\end{document}